\documentstyle[prb,aps,epsf,floats,twocolumn]{revtex}
\begin{document}
\draft  
%%%%%%%%%%%%%%%%%%%%%
\author{Yoshihiro Takushima, Akihisa Koga, and Norio Kawakami}
%%%%%%%%%%%%%%%%%%%%
\address{Department of Applied Physics, 
Osaka University, Suita, Osaka 565-0871, Japan}
%%%%%%%%%%%%%%%%%%%%%%%%%%%%%%%%%%%%%%%%%%
%%\title{Haldane gap system in a staggered field : mixed-spin model treatment}
\title{Magnetic Double Structure for $S=1, 1/2$ Mixed-Spin Systems}
%%%%%%%%%%%%%%%%%%%%%%%%%%%%%%%%%%%%%%%
\date{\today}
\sloppy
\maketitle
\begin{abstract}
We investigate the zero-temperature and the
finite-temperature properties of the two-dimensional 
antiferromagnetic quantum spin system composed of the 
$s=1/2$ and $s=1$ spins. The spin excitation spectrum 
as well as the thermodynamic quantities are computed 
by means of the Schwinger-boson mean-field theory.
%%%%%%%%%%%%%%%%%%%%%%%%%%%%%%%%%%%%%%%%%%%%%%%%%%%%%%%%%%%%%%%%%%%%%%
%            Revised 1                                                         
%%%%%%%%%%%%%%%%%%%%%%%%%%%%%%%%%%%%%%%%%%%%%%%%%%%%%%%%%%%%%%%%%%%%%%
We discuss how the magnetic double structure with the gapful and the 
gapless excitations is generated, and apply the results systematically 
to the Haldane gap system in a staggered magnetic field and also to the 
weakly coupled ferrimagnetic mixed-spin chains. It is confirmed 
that the results obtained  are  consistent with the experiments for 
the quasi-one-dimensional Haldane compounds $R_2 \rm BaNiO_5$.
%%%%%%%%%%%%%%%%%%%%%%%%%%%%%%%%%%%%%%%%%%%%%%%%%%%%%%%%%%%%%%%%%%%%%%
%By observing how the magnetic double structure with the gapful and 
%the gapless excitations generates,  we show that 
%the Haldane gap system in a staggered magnetic field
%is adiabatically connected to the mixed-spin chain with 
%ferrimagnetic ground state. It is confirmed that the results 
%obtained  are  consistent with the experiments for 
%the quasi-one-dimensional Haldane compounds $R_2 \rm BaNiO_5$.
%%%%%%%%%%%%%%%%%%%%%%%%%%%%%%%%%%%%%%%%%%%%%%%%%%%%%%%%%%%%%%%%%%%%%%

\end{abstract}
%\begin{multicols}{2}

%%%%%%%%%%%%%%%%%%%%%%%%%%%%%%%%%%%%%%%%%%%%%%%%%%%%%%%%%%%%%%%%%%%%%%
\section{Introduction}%%%%%%%%%%%%%%%%%%%%%%%%%%%%%%%%%%%%%%%%%%%%%%%%
%%%%%%%%%%%%%%%%%%%%%%%%%%%%%%%%%%%%%%%%%%%%%%%%%%%%%%%%%%%%%%%%%%%%%%
The Haldane gap system\cite{Haldane} 
has been one of the most fascinating subjects
in condensed matter physics, for which 
extensive experimental and theoretical investigations have been
providing a variety of new interesting phenomena. 
One of the hot topics is the magnetic double structure 
in the quasi-one-dimensional (1D)  Haldane systems
observed for the rare-earth compounds $R_2 \rm BaNiO_5$.
\cite{Hal-ch1,Nd-E1,Nd-E2,longi,cond,Pr}
In the case with $R=\rm Y^{3+}$, the system has a disordered ground state 
with the Haldane gap. \cite{HalGap1}
If $R$ is substituted by other magnetic ions such as $\rm Nd^{3+}$
and $\rm Pr^{3+}$,\cite{Pr,Str} the $s=1/2$ spins on the 
$R^{3+}$ sublattice order magnetically at low temperatures, thus 
giving rise to the magnetic double structure composed of 
the Haldane gap excitation and
the gapless excitation induced by the long-range order.
This system has been theoretically treated as the 
Haldane chains with  a  {\it static}
staggered field which are induced by the ordered  $s=1/2$ spins 
on the $R^{3+}$ sublattice so far.\cite{Gin,HcDMRG}
In a recent paper,\cite{longi} however, it has  been 
pointed out that the dynamics of the $s=1/2$ spins may be also 
important, stimulating us to treat the system by a mixed-spin model.
Concerning the magnetic double structure, there is another 
interesting  spin system of current interest, i.e. the 
quantum ferrimagnetic chain composed of 
 two kinds of mixed spins.\cite{ferri,Yee,Pati,Yama,Yama4,Wu} 
Several compounds have been already found, which indeed realize 
the mixed-spin system.\cite{ferri,Yee} This has
been stimulating further intensive theoretical studies on this 
subject.\cite{Pati,Yama,Yama4,Wu}
In particular, it has been pointed out that such a  ferrimagnetic  chain
has the double structure for the spin excitation spectrum,
which controls the characteristic properties in the
ferrimagnetic chain.\cite{Yama4}

The above two subjects, which have been studied experimentally 
in different contexts, should possess the common interesting 
physics, because both systems are characterized by the mixture of
two kinds of distinct spins.
Motivated by these hot topics, we here study the properties of 
the antiferromagnetic mixed-spin system in detail, 
for which the $s=1$ and $s=1/2$
spin chains are stacked alternately. We exploit the
two-dimensional (2D) system as the simplest model 
which possesses the magnetic long-range order at zero temperature. 
%%%%% Revise 1-1 %%%%%%%%%%%%%%%%%%%%%%%%%%%%%%
%%We shall explicitly  demonstrate that the 
%%Haldane gap system in a staggered field has the important connection 
%%to the  ferrimagnetic chain.
%%%%%%%%%%%%%%%%%%%%%%%%%%%%%%%%%%%%%%%%%%%%%
By computing the dispersion relation and the staggered magnetization 
by means of the Schwinger-boson mean-field theory,\cite{DPAro}
we clarify how our system generates the magnetic double structure,
which naturally interpolates the above two interesting spin systems.
We also show that the results obtained 
are consistent with the experimental findings
for quasi-1D Haldane compounds
$R_2 \rm BaNiO_5$.\cite{Hal-ch1,Nd-E1,Nd-E2,longi,cond,Pr} 

This paper is organized as follows.  We introduce the model
and then  briefly summarize the Schwinger-boson techniques
in Sec. II.  We discuss the zero-temperature properties in Sec. III, 
and finally move to the thermodynamic properties in Sec. IV.
A brief summary is given in Sec. V.

%%%%%%%%%%%%%%%%%%%%%%%%%%%%%%%%%%%%%%%%%%%%%%%%%%%%%%%%%%%%%%%%%%%%%%
\section{Schwinger-boson Mean-Field Theory}
%%\section{Model Hamiltonian}%%%%%%%%%%%%%%%%%%%%%%%%%%%%%%%%%%%%%%%%%%%
%%%%%%%%%%%%%%%%%%%%%%%%%%%%%%%%%%%%%%%%%%%%%%%%%%%%%%%%%%%%%%%%%%%%%%

Let us consider a 2D mixed-spin model on the square lattice,
which is described by the following Hamiltonian,
%%%%%%%%%%%%%%%%%%%%%%%%%%%%%%%%%%%
\begin{eqnarray}
H&=&J_{1}\sum_{i,j,\eta}\left[{\bf S}_{2i,2j}^{A_{1}}
\cdot{\bf S}_{2i+\eta,2j}^{B_{1}}+
{\bf S}_{2i+\eta,2j+1}^{B_{2}}\cdot{\bf S}_{2i,2j+1}^{A_{2}}
\right]\nonumber\\
&+&\sum_{i,j,\eta}\left[J_{2}{\bf S}_{2i,2j}^{A_{1}}
\cdot{\bf S}_{2i,2j+\eta}^{A_{2}}+
J_{3}{\bf S}_{2i+1,2j}^{B_{1}}\cdot{\bf S}_{2i+1,2j+\eta}^{B_{2}}\right],
\label{eq:model}
\end{eqnarray}
%%%%%%%%%%%%%%%%%%%%%%%%%%%%%%%%
where ${\bf S}_{i,j}^{A}$ (${\bf S}_{i,j}^{B}$) is the spin operator 
at the $(i,j)$-th site in the $(x,y)$ plane, and $\eta$ implies the 
summation to be taken over nearest-neighbor sites. 
All the exchange couplings  $J_{1}, J_{2}$ and $J_{3}$
are assumed to be antiferromagnetic.
We here focus on the mixed-spin system 
composed of $S^{A}=1/2$ and $S^B=1$ spins, 
since it is straightforward to generalize the results 
to the arbitrary-spin case.
%%%%%%%%%%%%%%%%%%%%%%%%
\begin{figure}[htb]
\epsfxsize=7cm
\centerline{\epsfbox{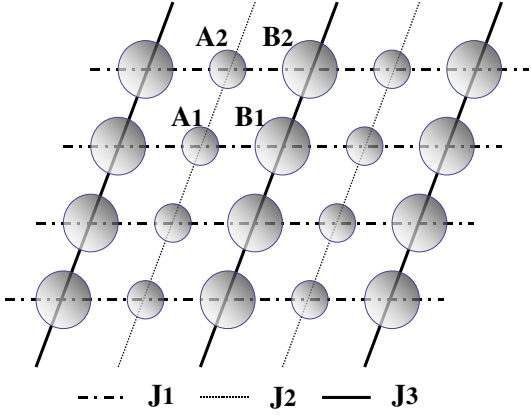}} 
\vspace{0.1cm}
\caption{2D mixed-spin model on a square lattice. 
The small (large) circle represents $S^A = 1/2$ ($S^B = 1$) spin.
The bold dashed, the thin dashed and the bold solid lines represent
the coupling constants $J_{1}, J_{2}$ and $J_{3}$, respectively.
$A_{1}, A_{2}, B_{1}$ and $B_{2}$ specify  four spins in the unit cell.}
\label{fig:md}%-----------------------------------------
\end{figure}
%%%%%%%%%%%%%%%%%%%%%%%%%
%%%%%%%%  Revised 2, FIG.1 J1,J2,J3 %%%%%
In Fig. \ref{fig:md}, we have drawn the mixed-spin model schematically.
The indices $A_m$ and $B_m$ label the sublattice in the unit cell.
Note that this mixed-spin model is constructed in two ways 
depending on how we stack the independent spin chains by
introducing three kinds of the coupling constants.
By taking $J_3\gg J_1, J_2$,
we can study the characteristic properties of the 
$s=1$ Haldane spin chains coupled with 
the $s=1/2$ gapless spin chains, which may have 
the relevance to  the Haldane gap system in a staggered field
observed in the compounds $R_2 \rm BaNiO_5$.
\cite{Hal-ch1,Nd-E1,Nd-E2,longi,cond,Pr}
On the other hand,  by setting the coupling 
constants $J_1\gg J_2, J_3$, we can investigate how the 
independent ferrimagnetic chains $(J_2=J_3=0)$, which have been studied 
extensively in recent years,\cite{ferri,Yee,Pati,Yama,Yama4,Wu}
are combined to form  the 2D system.
The advantage of our mixed-spin approach is that we can 
systematically describe these two 
interesting systems by continuously varying the parameters
 $J_1, J_2$ and $J_3$.

%%%%%%%%%%%%%%%%%%%%%%%%%%%%%%%%%%%%%%%%%%%%%%%%%%%%%%%%%%%%%%%%%
%%\section{Schwinger-boson Mean-Field Theory}
%%%%%%%%%%%%%%%%%%%%%%%%%%%%%%%%%%%%%%%%%%%%%%%%%%%%%%%%%%%%%%%%%

%%%%%%  Revised 3, ADDED REF. SWT %%%%%%%%%
We employ the Schwinger-boson mean-field theory (SBMFT)\cite{DPAro}
to study the above  mixed-spin model.\cite{SWT}
This method was applied successfully to the 2D Heisenberg 
model with uniform spins,\cite{DPAro,Sar}  and then  to
the bilayer system,\cite{bil} the double-exchange system,\cite{Sar1} 
the ferrimagnetic chain,\cite{Wu} etc.
It is known that  the SBMFT can describe the magnetically ordered phase,
which is characterized by the condensation of the 
Schwinger-bosons.\cite{Sar} In the Schwinger-boson representation, 
the spin operators are expressed in terms of the boson creation and
annihilation operators $\gamma_{\alpha}^{\dagger}$,  $\gamma_{\alpha}$,
with the Pauli matrix {\boldmath $\sigma$} as ${\bf S}=\frac{1}{2}
\gamma_{\alpha}^{\dagger}${\boldmath $\sigma$}$_{\alpha\beta}\gamma_{\beta}$ 
$(\alpha,\beta=\uparrow,\downarrow)$.
%%%%%%%%%%%%%%%%%%%%%%%%%%%%%%%
%%%%%%%%%%%%%%%%%%%%%%%%%%%%%%%%%%%%%%%%%%%%%%%%%%%%%%%%%%%%%%%%%%%%%%
Since the unit cell in our model includes four sites,
we necessarily introduce eight kinds of Bose operators 
$\gamma_{\alpha}=a^{(1)}_{\alpha}, a^{(2)}_{\alpha}, 
b^{(1)}_{\alpha}, b^{(2)}_{\alpha}$ %$(\alpha=\uparrow,\downarrow)$ 
which belong to the $A_1, A_2, B_1, B_2$ sublattices, respectively.
By imposing the constraint
$\gamma_{i\uparrow}^{\dagger}\gamma_{i\uparrow}+
\gamma_{i\downarrow}^{\dagger}\gamma_{i\downarrow} = 2S^{A}$ or $2S^{B}$
on each site, we can correctly map
the original spin system to the boson system.
Introducing the Lagrange multipliers 
$\lambda^{A}_{ij}$ and $\lambda^{B}_{ij}$, 
the Hamiltonian with the constraints is recast to 
%%%%%%%%%%%%%%%%%%%%%%%%%%%%%%%%%%%%%%%%%%%%%%%%%%%%%%%%
\begin{eqnarray}
H&=&-2\sum_{i,j,\eta}\left[J_{1}\left(A_{ij\eta}^{\dagger}A_{ij\eta}+
B_{ij\eta}^{\dagger}B_{ij\eta}\right)\right.\nonumber\\
&+&\left.J_{2}C_{ij\eta}^{\dagger}C_{ij\eta}+
J_{3}D_{ij\eta}^{\dagger}D_{ij\eta}\right]\nonumber\\
%%%%%%%%%%%%
&+&\sum_{n=1}^{2}\left[\sum_{(ij)\in A_n} \lambda^{A}_{ij}
\left(\sum_\sigma a_{ij\sigma}^{(n)\dagger}a_{ij\sigma}^{(n)}-
2S^{A}\right)\right.\nonumber\\
&+&\left.\sum_{(ij)\in B_n} \lambda^{B}_{ij}
\left(\sum_\sigma b_{ij\sigma}^{(n)\dagger}b_{ij\sigma}^{(n)}-
2S^{B}\right)\right]\nonumber\\
&+&\sum_{i,j,\eta}\left[
J_{1}\left(S^{A_{1}}S^{B_{1}}+S^{A_{2}}S^{B_{2}}\right)\right.\nonumber\\
&+&\left.
J_{2}S^{A_{1}}S^{A_{2}}+J_{3}S^{B_{1}}S^{B_{2}}\right],
\label{eq:hamil}
\end{eqnarray}
%%%%%%%%%%%%%%%%
where four bond operators are introduced as  
%%%%%%%%%%%%%%%%
\begin{eqnarray}
A_{ij\eta}&=&a^{(1)}_{2i,2j\uparrow}b^{(1)}_{2i+\eta,2j\downarrow}
-a^{(1)}_{2i,2j\downarrow}b^{(1)}_{2i+\eta,2j\uparrow},\\
B_{ij\eta}&=&a^{(2)}_{2i,2j+1\uparrow}b^{(2)}_{2i+\eta,2j+1\downarrow}
-a^{(2)}_{2i,2j+1\downarrow}b^{(2)}_{2i+\eta,2j+1\uparrow},\\
C_{ij\eta}&=&a^{(1)}_{2i,2j\uparrow}a^{(2)}_{2i,2j+\eta\downarrow}
-a^{(1)}_{2i,2j\downarrow}a^{(2)}_{2i,2j+\eta\uparrow},\\
D_{ij\eta}&=&b^{(1)}_{2i+1,2j\uparrow}b^{(2)}_{2i+1,2j+\eta\downarrow}
-b^{(1)}_{2i+1,2j\downarrow}b^{(2)}_{2i+1,2j+\eta\uparrow}.
\end{eqnarray}
%%%%%%%%%%%%%%%%%%%%%%%%%%%%%%%%%%%%%%%%%%%%%%%%%%%%%%
We perform a Hartree-Fock decomposition of eq. (\ref{eq:hamil}) 
by taking the thermal average $<A_{ij\eta}>=A$, 
$<\lambda^{A}_{ij}>=\lambda_{A}$, etc.,
which means that these values are assumed to be uniform and static.
%%%%%%%%%%%%%%%%%%%%%%%%%

%%%%%%%%%%%%%%%%%%%
By diagonalizing the mean-field Hamiltonian via the 
Bogoliubov transformation, we have 
%%%%%%%%%%%%%%%%%%%%%%%%%%%%%%%%%%%%%%
\begin{eqnarray}
H_{\rm MF}&=&\sum_{{\bf k}\sigma}\sum_{n=1}^{2}\left[
E_{\bf k}^{(1)}\alpha_{{\bf k}\sigma}^{(n)\dag}
\alpha_{{\bf k}\sigma}^{(n)}+
E_{\bf k}^{(2)}\beta_{{\bf k}\sigma}^{(n)\dag}
\beta_{{\bf k}\sigma}^{(n)}\right],
\label{bogo}
\end{eqnarray}
%%%%%%%%%%%%%%%%%%%%%%%%%%%%%%%%%%%%%
where $\alpha$ and $\beta$ are the Bose operators for 
normal modes.
The corresponding energy spectrums read
%%%%%%%%%%%%%%%%%%%%
\begin{eqnarray}
E_{\bf k}^{(1)}&=&\sqrt{\frac{E_{0}-\sqrt{E_{1}}}{2}},\ \ \ 
E_{\bf k}^{(2)}=\sqrt{\frac{E_{0}+\sqrt{E_{1}}}{2}}, 
\end{eqnarray}
%%%%%%%%%%%%%%%%%%%%
where $E_{0}$ and $E_{1}$ are given by
%%%%%%%%%%%%%%%%%%%%
\begin{eqnarray}
E_{0}&=&\lambda_{A}^{2}+\lambda_{B}^{2}-2d_{\bf k}^{2}
-e_{\bf k}^{2}-f_{\bf k}^{2},
\nonumber\\
&&\\
E_{1}&=&(\lambda_{A}^{2}-\lambda_{B}^{2}-e_{\bf k}^{2}
+f_{\bf k}^{2})^{2}\nonumber\\
&&\nonumber\\
&&-4d_{\bf k}^{2}\left( (\lambda_{A}-\lambda_{B})^{2}
-(e_{\bf k}+f_{\bf k})^{2}\right),
\nonumber
\end{eqnarray}
%%%%%%%%%
with
%%%%%%%%%
\begin{eqnarray}
d_{\bf k}&=&2AJ_{1}\cos k_{x},\ \ \ e_{\bf k}=2CJ_{2}\cos k_{y},\nonumber\\
f_{\bf k}&=&2DJ_{3}\cos k_{y}.\nonumber
\end{eqnarray}
%%%%%%%%%%%%%%%%%%%%%%%%%
By minimizing the free energy thus obtained at finite temperatures,
we  end up with  the self-consistent equations
for $A=B, C, D, \lambda_{A}, \lambda_{B}$, 
%%%%%%%%%%%%%%%%%%%%%%%%
\begin{eqnarray}
&&1+2S^{A}=
\sum_{n}\int \frac{d{\bf k}}{\pi^{2}}
\coth \kappa E_{\bf k}^{(n)} \frac{\partial E_{\bf k}^{(n)}}
{\partial\lambda_{A}},
\label{sel1}
\\
&&1+2S^{B}=
\sum_{n}\int \frac{d{\bf k}}{\pi^{2}}
\coth \kappa E_{\bf k}^{(n)} \frac{\partial E_{\bf k}^{(n)}}
{\partial\lambda_{B}},
\label{sel2}
\\
&&-8J_{1}A=
\sum_{n}\int \frac{d{\bf k}}{\pi^{2}}
\coth \kappa E_{\bf k}^{(n)} \frac{\partial E_{\bf k}^{(n)}}{\partial A},
\label{sel3}\\
&&-4J_{2}C=
\sum_{n}\int \frac{d{\bf k}}{\pi^{2}}
\coth \kappa E_{\bf k}^{(n)} \frac{\partial E_{\bf k}^{(n)}}{\partial C},
\label{sel4}\\
&&-4J_{3}D=
\sum_{n}\int \frac{d{\bf k}}{\pi^{2}}
\coth \kappa E_{\bf k}^{(n)} \frac{\partial E_{\bf k}^{(n)}}{\partial D},
\label{sel5}
\end{eqnarray}
%%%%%%%%%%%%%%%%%%%%%%%%
with $\kappa=1/(2k_BT)$,
where we have assumed that the bond operators which link the $s=1$ 
and $s=1/2$ spins take the same mean value, $A = B$.
This completes our formulation based on the SBMFT.  In the following
sections, we solve  these self-consistent equations to estimate the 
excitation spectrum and the thermodynamic quantities. 
Since it is not easy to analytically perform the Bogoliubov transformation,
we  numerically diagonalize the mean-field Hamiltonian
to compute the energy dispersions $E_{\bf k}^{(1)}$ and $E_{\bf k}^{(2)}$.

%%%%%%%%%%%%%%%%%%%%%%%%%%%%%%%%%%%%%%%%%%
\section{Properties at Zero Temperature}
%%%%%%%%%%%%%%%%%%%%%%%%%%%%%%%%%%%%%%%%%%%%%%%%%

In order to treat the ground state properties, it should be 
taken into account that the Bogoliubov particles of the $\alpha$ branch
in eq.(\ref{bogo})
may  condense at absolute zero, because
the excitation energy $E^{(1)}_{\bf k}$ 
has its minimal value $E^{(1)}_{\bf k}=0$ at ${\bf k=0}$
 while the $\beta$ branch has a finite gap even at $T=0$. 
Sarker {\it et al.}\cite{Sar} showed that the long-range order 
is described by  the condensation of the Schwinger-bosons for   
the ferromagnetic and antiferromagnetic Heisenberg models. 
This is also the case for our 2D mixed-spin model on a square lattice. 

Suppose that the bosons condense at the states 
of $\alpha^{(1)}_{\uparrow}|_{\bf k=0}$ and 
$\alpha^{(2)}_{\downarrow}|_{\bf k=0}$, by fictitiously 
applying an infinitesimal external staggered field to the
$A$ and $B$ lattices.  The self-consistent equations at 
$T=0$ , which include the Bose condensation,
now read,
%%%%%%%%%%%%%%%%%%%%%%%%%%%%%%%%%%%%%%%
\begin{eqnarray}
\Gamma&=&\frac{2}{N^{2}}
\coth \kappa E_{\bf k=0}^{(1)}|_{\kappa\rightarrow\infty}
\nonumber\\
&=&\frac{\left[1+2S^{A}
-\sum_{n}\int \frac{d{\bf k}}{\pi^{2}}
\frac{\partial E_{\bf k}^{(n)}}{\partial \lambda_{A}}\right]}
{\partial E_{\bf k=0}^{(1)}/\partial \lambda_{A}},\\
1+2S^{B}
&=&\Gamma\left.
\frac{\partial E_{\bf k}^{(1)}}{\partial\lambda_{B}}\right|_{\bf k=0}
+\sum_{n}\int\frac{d{\bf k}}{\pi^{2}}\frac{\partial E_{\bf k}^{(n)}}
{\partial \lambda_{B}},\\
-4A&=&\Gamma\left.
\frac{\partial E_{\bf k}^{(1)}}{\partial d_{\bf k}}\right|_{\bf k=0}
+\sum_{n}\int\frac{d{\bf k}}{\pi^{2}}\cos k_{x}
\frac{\partial E_{\bf k}^{(n)}}
{\partial d_{\bf k}},\nonumber\\
&&\\
-2C&=&\Gamma\left.
\frac{\partial E_{\bf k}^{(1)}}{\partial e_{\bf k}}\right|_{\bf k=0}
+\sum_{n}\int\frac{d{\bf k}}{\pi^{2}}\cos k_{y}
\frac{\partial E_{\bf k}^{(n)}}
{\partial e_{\bf k}},\nonumber\\
&&\\
-2D&=&\Gamma\left.
\frac{\partial E_{\bf k}^{(1)}}{\partial f_{\bf k}}\right|_{k=0}
+\sum_{n}\int\frac{d{\bf k}}{\pi^{2}}\cos k_{y}
\frac{\partial E_{\bf k}^{(n)}}
{\partial f_{\bf k}}.\nonumber\\
&&
\end{eqnarray}
%%%%%%%%%%%%%%%%%%%%%%%%%%%%%%%%%%%%%%%
We shall solve these equations numerically for given coupling 
constants $J_{1}, J_{2}$ and $J_{3}$.

%%%%%%%%%%%%%%%%%%%%%%%%%%%%%%%%%
\subsection{Dispersion relation}
%%%%%%%%%%%%%%%%%%%%%%%%%%%%%%%%%%

\begin{figure}[htb]
\epsfxsize=8cm
\centerline{\epsfbox{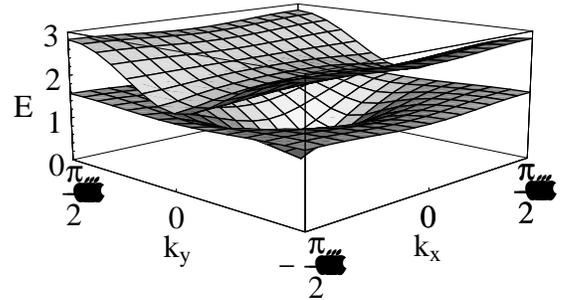}} 
\vspace{0.1cm}
\caption{The excitation spectrum for the mixed-spin model 
which may be regarded as the {\it Haldane gap in a staggered field}
($J_{1}=J_{2}=1/2 ,J_{3}=1$).  
Note that the Brillouin zone is reduced to a quarter of 
that for the uniform spin case,
since the unit cell now includes the four sites.
}
\label{fig:sp2}%-----------------------------------------
\end{figure}

We start our discussions with the case that may be regarded as
the Haldane gap system in a staggered field. When $J_1=0$, the 2D 
system is completely decoupled into the $s=1$ massive Haldane
chains and the $s=1/2$ massless spin chains. When we introduce the 
interchain couplings among them, the ground state has the long-range 
order mainly due to the $s=1/2$ spins. The important point is that 
even though we have the long-range order, the Haldane-type
gapful excitation still exists.  Therefore, as far as  
the case with small $J_1$ and $J_2$ is concerned, the system
is regarded as  the one often called the Haldane gap system 
in a staggered field.\cite{Hal-ch1,Nd-E1,Nd-E2,longi,cond,Pr}
In Fig. $\ref{fig:sp2}$, we show the dispersion relation
calculated for $J_{1}=J_{2}=1/2$ and $J_{3}=1$.
%%%%%%  MOVED TO BELOW %%%%%%%%%%%%%%
%%Although we have also performed 
%%the calculation for the cases with 
%%smaller $J_1$ and $J_2$,  the obtained dispersion of the 
%%optical branch becomes almost flat in the $k_x$-direction,
%%so that we have not shown them here.\cite{comment}
%%%%%%%%%%%%%%%%%%%%%%%%%%%%%%%%%%%%%%%%%%%%%%%%%%
%%%%%%%%%%%%%%%%%%%%%%%%%%%%%%%%%%%%%%%%%%%
As seen in this figure, the lower branch is gapless  with linear 
dispersion relation, reflecting the antiferromagnetic long-range order,
whereas  the upper optical mode is mainly 
 composed of the Haldane-type excitation. 
It is observed  that the dispersion in the $k_{x}$-direction is indeed weak
for the optical mode, since the interchain coupling is much smaller than the 
energy gap for the optical mode, 
which confirms that the system in Fig. \ref{fig:sp2}
may be regarded as the Haldane gap system in a staggered field.  
%%%%%%%  Revised 4   %%%%%%%%
%%%%%%%  MOVED FROM ABOVE  %%%%%%%%%%%%%%%%%%%%%%%%%%%
Although we have also performed 
the calculation for the cases with 
smaller $J_1$ and $J_2$,  the obtained dispersion of the 
optical branch becomes almost flat in the $k_x$-direction,
 so that we have not shown them here.
%%%%%  PREVIOUS REFERENCE, MOFDIFIED  %%%%%%%
Concerning this limit of small $J_1$, we here make a brief comment
on the validity of the SBMFT. When $J_1$ takes too small value, 
for which the system is almost decoupled into independent $s=1$ and 
$s=1/2$ chains, the SBMFT may lead to a pathological result: although
the Haldane gap for the $s=1$ chain is well described by the SBMFT, 
it is not the case for the gapless $s=1/2$ chains, for which we are 
left with a gapful phase. Actually, if the value of $J_1$ ($=J_2$) becomes 
smaller than $J_1/J_3 \sim 10^{-2}$, we encounter a problem that the 
present self-consistent calculation does not converge, implying 
that our assumption for the antiferromagnetic ordered state does not 
hold anymore. Nevertheless, we find that the correct behavior with 
the antiferromagnetic ground state is still obtained except for this 
small parameter region.
%%%%%%%%%%%%%%%%%%%%%%%%%%%%%%%%%%%%%%%%%%%%%%%%%%%%%%%%%%%%%%%%%

\begin{figure}[htb]
\epsfxsize=8cm
\centerline{\epsfbox{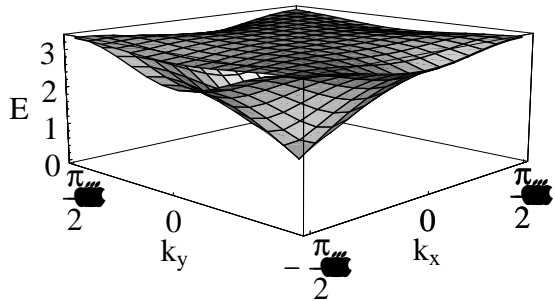}} 
\vspace{0.1cm}
\caption{The excitation spectrum for the 2D mixed-spin model 
($J_{1}=J_{2}=J_{3}=1$).}
\label{fig:sp1}%-----------------------------------------
\end{figure}

\begin{figure}[htb]
\epsfxsize=8cm
\centerline{\epsfbox{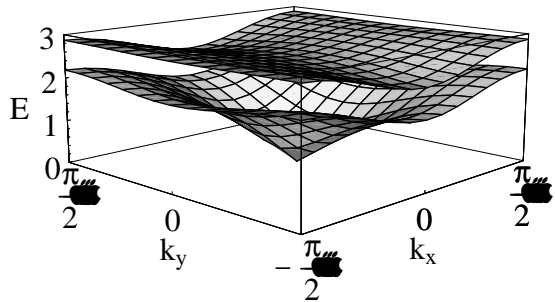}} 
\vspace{0.1cm}
\caption{The excitation spectrum for the {\it coupled ferrimagnetic chains}
($J_{1}=1 ,J_{2}=J_{3}=1/2$).}
\label{fig:sp3}%-----------------------------------------
\end{figure}

By  increasing  the coupling parameters $J_{2}, J_{3}$  
continuously, we naturally enter 
in the 2D ordered mixed-spin system. Note that although during this process
the magnetic double structure is kept unchanged
in its typical feature, the nature of the optical mode
is gradually changed from the Haldane-gap excitation: i.e.
the gapful excitation may be equally contributed from 
 both the $s=1$ and $s=1/2$ spin sectors.
 As a reference, we show the dispersion relation
for the mixed-spin model with the isotropic bonds $J_1=J_2=J_3=1$ 
in Fig. \ref{fig:sp1}.
%%%%%%%%%%%%%%%%%%%%%%%%%%%%%%%%%%%%%%%%%%%%%%%%%%
%%%%%%%%%%%%%%%%%%%%%%%%%%%%%%%%%%%%%%%%%%%
If we further decrease the couplings $J_{2}$ and $J_{3}$, 
the system gradually approaches the quasi-1D 
ferrimagnetic chains with the periodic arrangement 
of spins $1/2 \circ 1 \circ 1/2 \circ 1$.
The  dispersion relation obtained in the
corresponding  parameter region
is shown in Fig. \ref{fig:sp3}.
%%%%%%%%%%%%%%%%%%%%%%%%%%%%%%%%%%%%%%%%%%%%%%%%%%
 %%%%%%%%%%%%%%%%%%%%%%%%%%%%%%%%%%%%%%%%%%%
Now the optical mode with a weak dispersion in the $k_{y}$-direction
 is essentially the same as that found for the 
ferrimagnetic chain.\cite{Yama4} 
%%%%%%%%%%%%%%%%%%%%%%%%%%%%%%%%%%%%%%%%%%%%%
%%　　　　　　　　　　　　　　　　　　　 　%%
%%  Revised 5, ADDED FIG, 
%%                                         %%
%%%%%%%%%%%%%%%%%%%%%%%%%%%%%%%%%%%%%%%%%%%%%
%%%%%%%%%%%%%%%%%%%%%%%%%%%%%%%%%%%%%%%%%%%%%%%%%%
\begin{figure}[htb]
\epsfxsize=8cm
\centerline{\epsfbox{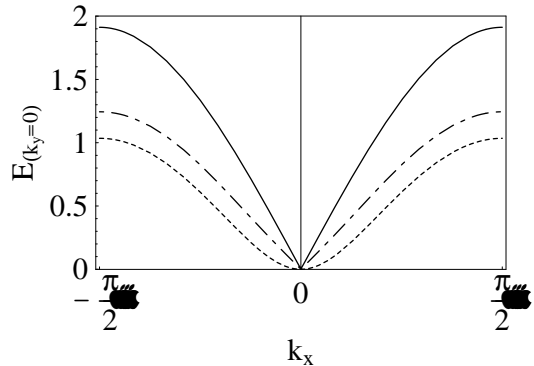}} 
\vspace{0.1cm}
\caption{The gapless dispersion for the 
{\it coupled ferrimagnetic chains}
 at $k_y$=0. The solid, the dash-dotted and the dashed lines  
correspond to the cases of  $J_2=J_3=0.5$, 0.1 and 0 ($J_1=1$).}
\label{fig:sp4}%-----------------------------------------
\end{figure}
%%%%%%%%%%%%%%%%%%%%%%%%%%%%%%%%%%%%%%%%%%%
%%%%%  Revised 6, FERRI LIMIT  kx^2 DEPENDENCE   %
%%%%%%%%%%%%%%%%%%%%%%%%%%%%%%%%%%%%%%%
In order to clearly observe how the gapless dispersion changes its 
character in the ferrimagnetic-chain limit, we have shown the low-energy
dispersion relation at $k_y$=0 in Fig. \ref{fig:sp4}. It is seen that 
with the decrease of the coupling constants $J_2$ and $J_3$, the 
$k_x$-linear dependence is gradually changed to the $k_x^2$ dependence 
characteristic of the ferrimagnetic chain except for the small $k_x$ 
region where the 2D antiferromagnetic order still gives rise to the 
$k_x$-linear dependence.  At $J_2=J_3=0$, the system is reduced to 
the isolated ferrimagnetic chains, for which the 2D character completely
disappears, and thus the gapless dispersion simply follows the 
$k_x^2$ dependence. 
%%%%%%%%%%%%%%%%%%%%%%%%%%%%%%%%%%%%%%%%%%%%%%%%%%%%%%%%%
%%%%%%%%%%%%%%%%%%%%%%%%%%%%%%%%%%%%%%%%%%%%%
%Actually, we have confirmed that 
%at $J_2=J_3=0$ the system is correctly reduced to the isolated 
%ferrimagnetic chains, for which the gapless
%dispersion shows the  $k_x^2$-dependence instead of the 
%$k_x$-linear dependence for the 2D case. 
%%%%%%%%%%%%%%%%%%%%%%%%%%%%%%%
We note that the ferrimagnetic chains  have been experimentally 
realized in the  compounds such as $\rm NiCu(C_2O_4)_2\cdot 4H_2O$ 
and $\rm MnCu(pba)(H_2O)_3\cdot 2H_2O$,\cite{Yee} and have been 
studied theoretically by many groups.\cite{Pati,Yama,Yama4} 
Among others, it has been reported \cite{Yama4} that this system 
exhibits the dual properties consistent with our
results: the physical quantities are controlled 
by the $s=1$ antiferromagnetic spin chain  at high temperatures 
and by  the effective $s=1/2$ ferromagnetic spin chain 
at low temperatures.

%%%%%%%%%%%%%%%%%%%%%%%%%%%%%%%%%%%%%%%%%%%%%%%%%%%%%%%%%%%%%%%%%%%%%%
%            Revised 7, ADDIABATICALLY      
%%%%%%%%%%%%%%%%%%%%%%%%%%%%%%%%%%%%%%%%%%%%%%%%%%%%%%%%%%%%%%%%%%%%%%
The above analysis of the excitation spectrum implies that 
the quasi-1D Haldane gap system in a staggered 
field\cite{Hal-ch1,Nd-E1,Nd-E2,longi,cond,Pr} and
the quasi-1D weakly coupled ferrimagnetic chains,\cite{ferri,Yee}
which have been studied in different contexts 
experimentally, share common interesting physics
inherent in the mixed-spin systems.  In particular, it is highly 
desirable to experimentally observe the magnetic double structure
in the excitation spectrum for the ferrimagnetic-chain  compounds.
%%%%%%%%%%%%%%%%%%%%%%%%%%%%%%%%%%%%%%%%%%%%%%%%%%%%%%%%%%%%%%%%%%%%%%
%The above analysis of the excitation spectrum implies that 
%the Haldane gap system in a staggered field and
%the ferrimagnetic chains, which have been studied in 
%different contexts experimentally,\cite{ferri,Yee}
%are adiabatically connected with  each other.
%We can thus say that the double structure observed 
%in the rare-earth compounds $R_2 \rm BaNiO_{5}$ 
%\cite{Hal-ch1,Nd-E1,Nd-E2,longi,cond,Pr}
%has essentially  the same 
%origin as in the ferrimagnetic chain.\cite{Yama4} 
%%%%%%%%

%%%%%%%%%%%%%%%%%%%%%%%%%%%%%%%%%%%%%%%%%%%
\subsection{Haldane gap in a staggered field} 
%%%%%%%%%%%%%%%%%%%%%%%%%%%%%%%%%%%%%%%%%%%

\begin{figure}[htb]
\epsfxsize=7cm
\centerline{\epsfbox{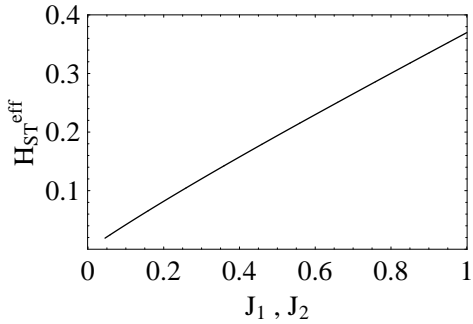}} 
\vspace{0.1cm}
\caption{The effective staggered field $H_{\rm ST}^{{\rm eff}}$
 as a function of 
the coupling constant $J_1$ (=$J_2$) where we set  $J_3=1$.}
\label{fig:hal2}%-----------------------------------------
\end{figure}

Let us  discuss  the case of the Haldane gap
system in a staggered field  in more detail.
We here observe  how the effective staggered field 
induced by the $s=1/2$ spin chains 
affects the properties  of the Haldane gap,
by changing the coupling constants $J_1$ and $J_2$.
For small values of $J_1$ and $J_2$, we define the 
effective staggered field on the $s=1$  Haldane chain
by $H_{\rm ST}^{\rm eff}=J_{1} <S_{z}^{A}>$, 
%%%%%%%%%%%%%%%%%%%%%%%%%%%%%%%
where $<S_{z}^{A}>$ 
is the spontaneous staggered magnetization of the $s=1/2$ spin chain.
%%%%%%%%%%%%%%%%%%%%%%
We numerically estimate the effective staggered field, 
and show  the obtained results in Fig. \ref{fig:hal2}.
%%%%%%%%%%%%%%%%%%%%%%%%%%%%%%%%%%%%%%%%%%%%%
%%%%%%%%%%%%%%%%%%%%%%%%%%%%%%%%%%%%%%%%%%%%%
It is seen that the effective staggered field
increases monotonously  with the increase of $J_1$ and $J_2$, 
as should be expected. In a similar way, we also compute the  
staggered magnetization  $<S_z^B>$ of the $s=1$ spin chains,
and as well as the Haldane gap which 
is defined as the minimum of the excitation energy
in the optical branch $\Delta=E_{{\bf k}=(\frac{\pi}{2},0)}^{(2)}$.
%%%%%%%%%%%%%%%%%%%%%%%%%%%%%%%%%%%%%%%%%%%%%
\begin{figure}[htb]
\epsfxsize=7cm
\centerline{\epsfbox{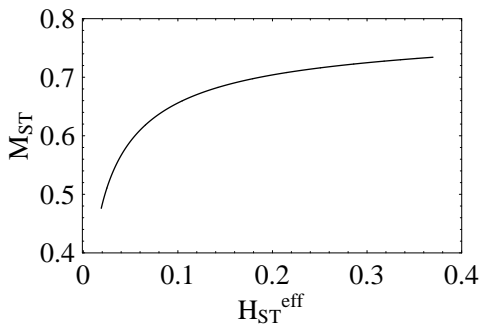}} 
\vspace{0.1cm}
\caption{Plots of the
staggered magnetization $M_{\rm ST }$ 
as a function of the effective staggered magnetic field.
}
\label{fig:hal1}%-----------------------------------------
\end{figure}
%%%%%%%%%%%%%%%%%%%%%%%%%%%%%%%%%%%%%%%%%%%%%
%%%%%%%%%%%%%%%%%%%%%%%%%%%%%%%%%%%%%%%%%%%%%
\begin{figure}[htb]
\epsfxsize=7cm
\centerline{\epsfbox{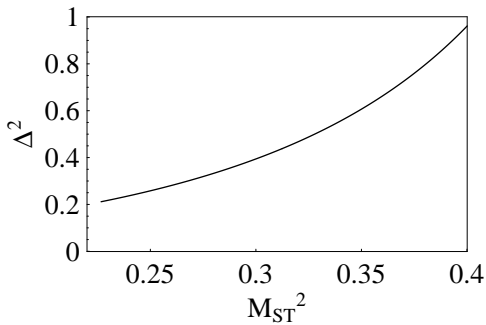}} 
\vspace{0.1cm}
\caption{Plots of the square of the Haldane gap $\Delta$ 
as a function of the square of the
staggered magnetization $M_{\rm ST }$.}
\label{fig:hal3}%-----------------------------------------
\end{figure}
%%%%%%%%%%%%%%%%%%%%%%%%%%%%%%%%%%%%%%%%%%%%%
These quantities are shown in Figs. \ref{fig:hal1} and \ref{fig:hal3}.
Following the way used for the experimental analysis,
\cite{Hal-ch1,Nd-E2}
we have plotted the staggered magnetization (the square of the 
Haldane gap) as a function of the 
effective staggered field (the square of the staggered 
magnetization). It is seen that as the staggered field is increased,
the staggered magnetization  as well as the Haldane 
gap $\Delta$  are increased, being consistent with 
the result pointed out for the Haldane chain 
system with a {\it static} magnetic field.
\cite{Gin,HcDMRG}

Since we are dealing with the 2D model at zero temperature, our 
results may not be directly applied to the experiments.
Nevertheless, 
we can confirm whether the present  results are consistent  with 
the experiments for the rare-earth compounds $R_2 \rm BaNiO_{5}$
\cite{Hal-ch1,Nd-E1,Nd-E2,longi,cond,Pr} with 
$R= \rm Nd^{3+}$ or $\rm Pr^{3+}$.\cite{Pr,Str}
Experimentally, as the temperature is decreased, the system shows the 
phase transition to the magnetically ordered phase. 
When the temperature is further decreased, the 
staggered moment for the  $s=1/2$ sector develops, 
and thereby gives rise to the increase of the  staggered field,
which indeed enhances  the 
magnitude of the Haldane gap.\cite{Gin,HcDMRG}  Also, the 
staggered moment on the Haldane chains increases, as 
should be expected. These 
characteristic features are consistent with our results, and 
in particular, the qualitative behaviors for the 
staggered magnetization and the Haldane gap  shown in
Figs. \ref{fig:hal1} and \ref{fig:hal3} agree fairly well
with experimental findings.\cite{Hal-ch1,Nd-E2}

%%%%%%%%%%%%%%%%%%%%%%%%%%%%%%%%%%%%%%%%%%%
\subsection{Coupled ferrimagnetic chains}
%%%%%%%%%%%%%%%%%%%%%%%%%%%%%%%%%%%%%%%%%%%%

\begin{figure}[htb]
\epsfxsize=7cm
\centerline{\epsfbox{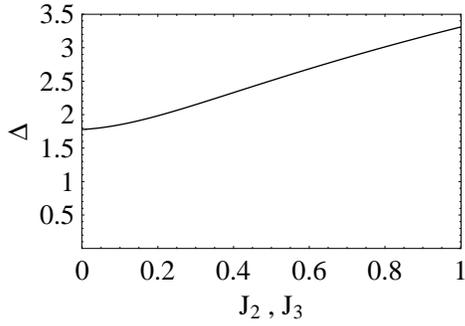}} 
\vspace{0.1cm}
\caption{Plots of the spin gap for the optical branch in the limit of 
the coupled ferrimagnetic chains ($J_2=J_3$).
}
\label{fig:FgapT0-1}%-----------------------------------------
\end{figure}

Before closing this section, we briefly discuss the case 
close to the ferrimagnetic
chain, which is realized by taking the limit of $J_1 \gg J_2, J_3$.
%%%%%%%%%%%%%%%%%%%%%%%%%%%%%%%%%%%%%%%%%%%%%%%%%%%%%%%%%%%%%%%%%%%%%%
%            Revised 8, FERIRI LIMIT      
%%%%%%%%%%%%%%%%%%%%%%%%%%%%%%%%%%%%%%%%%%%%%%%%%%%%%%%%%%%%%%%%%%%%%%
Even in  this one-dimensional limit, the system 
still exhibits the antiferromagnetic order as far as 
$J_2$ and $J_3$ take finite values.  As mentioned above,
at $J_2=J_3=0$, the system shows the ferrimagnetic order, and hence 
the dispersion relation for the acoustic mode is changed to the quadratic 
one.\cite{Yama,Yama4,Wu}  In this way,
the crossover-like behavior is seen in the gapless mode, whereas
 much simpler behavior is observed in the gapful mode. 
%%%%%%%%%%%%%%%%%%%%%%%%%%%%%%%
%Even in  this one-dimensional limit of independent chains, the system 
%still exhibits the ferrimagnetic order as is the case for the 
%ferromagnetic chain.  Hence, the dispersion relation for the
%acoustic mode has the quadratic dependence on the momentum,
%as mentioned above.\cite{Yama,Yama4,Wu}  
%%%%%%%%%%%%%%%%%%%%%%%%%%%%%%%%%%%%%%%%%%%%%%%%%%%%%%%%%%%%%%%
%            Revised 8, COMPARISON WITH QMC
%%%%%%%%%%%%%%%%%%%%%%%%%%%%%%%%%%%%%%%%%%%%%%%%%%%%%%%%%%%%%%%  
In Fig. \ref{fig:FgapT0-1}, we display 
the spin gap $\Delta$ for the optical branch in the ferrimagnetic-chain
limit. With decreasing $J_2$ and $J_3$, the spin gap monotonically 
decreases, and reaches the value of  $\Delta$=1.778 at $J_{2}=J_{3}=0$, 
which is very close to that of the quantum Monte Carlo method, 
$\Delta$=1.767,\cite{Yama}
as already demonstrated by Wu {\it et al}.\cite{Wu}  In this way, in the 
ferrimagnetic-chain limit, the SBMFT may provide the reliable 
estimates of the physical quantities  even at quantitative level.
%%%%%%%%%%%%%%%%%%%%%%%%%%%%%%%%%%%%%%%%%%%%
As discussed by Yamamoto and Fukui,\cite{Yama4} this optical mode
is mainly composed of excitations in the 
effective $s=1$ antiferromagnetic spin chain, whereas the
acoustic mode is given by excitations in  
the effective $s=1/2$ ferromagnetic chain.
%%%%%%%%%%%%%%%%%%%%%%%%%%%%%%%%%%%%%%%%%%%%%
%Wu {\it et al}.\cite{Wu}
% have already studied this ferrimagnetic chain  by means
%of the Schwinger-boson method.  Our results indeed  reproduce 
%their results in this case. As a reference, 
%the spin gap for the optical branch is shown in Fig. \ref{fig:FgapT0-1}.
%%%%%%%%%%%%%%%%%%%%%%%%%%%%%%%%%%%%%%%%%%%%%
%%%%%%%%%%%%%%%%%%%%%%%%%%%%%%%%%%%%%%%%%%%%%
%We note that at $J_2=J_3=0$, the spin gap exactly 
%coincides with that obtained by Wu {\it et al}.\cite{Wu}
%As discussed by Yamamoto and Fukui,\cite{Yama4} this optical mode
%is mainly composed of excitations in the 
%effective $s=1$ antiferromagnetic spin chain, whereas the
%acoustic mode is given by excitations in  
%the effective $s=1/2$ ferromagnetic chain.

%%%%%%%%%%%%%%%%%%%%%%%%%%%%%%%%%%%%%%%%%%%%%%%%%%%%%%%%%%%%%%%%%
\section{Thermodynamic Properties at Finite Temperatures}
%%%%%%%%%%%%%%%%%%%%%%%%%%%%%%%%%%%%%%%%%%%%%%%%%%%%%%%%%%%%%%%%%
We now move to the thermodynamic properties at finite temperatures. 
The self-consistent equations (\ref{sel1})-(\ref{sel5}) are solved
numerically at finite temperatures to obtain the 
thermodynamic quantities.  We here briefly summarize the 
results obtained. In the previous section, it has been shown 
that even for the set of the parameters
 $J_{1}=J_{2}=1/2,J_{3}=1$ ($J_{1}=1,J_{2}=J_{3}=1/2$),
the system may be approximately
described by the Haldane gap system in a staggered field
(coupled ferrimagnetic chains), so that we 
shall show the results for these parameters below.
%%%%%%%%%%%%%%%%%%%%%%%%%%%%%%%%%%%%%%%%%%%%%
\begin{figure}[htb]
\epsfxsize=7cm
\centerline{\epsfbox{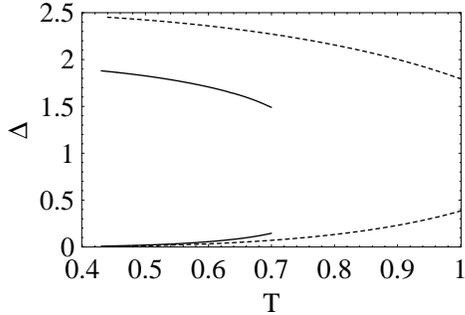}} 
\vspace{0.1cm}
\caption{The spin gaps for the optical  and 
acoustic branches as a function of the
temperature $T$: the solid lines correspond to 
the case for  $J_{1}=J_{2}=1/2,J_{3}=1$ while the dashed lines 
 for $J_{1}=1, J_{2}=J_{3}=1/2$. Note that at higher temperatures 
our SBMF approach breaks down, so that we have plotted the data 
available for each choice of parameters.}
\label{fig:half1}%-----------------------------------------
\end{figure}
%%%%%%%%%%%%%%%%%%%%%%%%%%%%%%%%%%%%%%%%%%%%%
We start with the effective spin gaps calculated at finite temperatures,
which are shown as a function of the 
temperature in Fig. \ref{fig:half1}.  It is seen that both of 
the two cases exhibit similar temperature dependence in the
spin gap.  For each case, there are two distinct 
spin gaps, corresponding respectively to
the optical mode and  the acoustic mode,
reflecting the double structure 
for the excitation spectrum. Note that 
both of two modes should be massive at finite temperatures.
It may be physically sensible to
regard these spin gaps as the inverse of the correlation lengths.
As should be expected, the spin gap for the optical mode increases
up to the zero-temperature value with the decrease of the temperature. 
On the other hand, the spin gap  for the 
acoustic mode decreases, leading to the divergent correlation length
which characterizes the  magnetically ordered state  at $T=0$.

The  uniform and staggered spin susceptibilities,
$\chi_{\rm uni}$ and $\chi_{\rm stag}$, are calculated 
by using the standard linear-response formulae,
%%%%%%%%%%%%%%%%%%%%%%%
\begin{eqnarray}
&&\chi_{\rm uni}=\sum_{i,j=1,2}\left[ 
\langle\langle S^{A_{i}}_{z}S^{A_{j}}_{z}\rangle\rangle_{{\bf q},\omega}
+\langle\langle S^{B_{i}}_{z}S^{B_{j}}_{z}\rangle\rangle_{{\bf q},\omega}
\right.\nonumber\\
&&\ \ \ \ \ \ \ \ \ \ \ \ +\left.
2\langle\langle S^{A_{i}}_{z}S^{B_{j}}_{z}\rangle\rangle_{{\bf q},\omega}
\right]|_{{\bf q},\omega\rightarrow 0},\\
\nonumber \\
&&\chi_{\rm stag}= \nonumber\\
&&\sum_{i,j=1,2}\left[(-1)^{i+j}( 
\langle\langle S^{A_{i}}_{z}S^{A_{j}}_{z}\rangle\rangle_{{\bf q},\omega}
+\langle\langle S^{B_{i}}_{z}S^{B_{j}}_{z}\rangle\rangle_{{\bf q},\omega}
)\right.\nonumber\\
&&\ \ \ \ \ \ \ \ \ \ \ \ +\left.2(-1)^{i+j+1}
\langle\langle S^{A_{i}}_{z}S^{B_{j}}_{z}\rangle\rangle_{{\bf q},\omega}
\right]|_{{\bf q},\omega\rightarrow 0},
\end{eqnarray}
%%%%%%%%%%%%%%%%%%%%%%%
where $\langle\langle S^{A_{1}}_{z}S^{A_{1}}_{z}\rangle\rangle
_{{\bf q},\omega}$, etc., 
are the retarded spin correlation functions. 
%%%%%%%%%%%%%%%%%%%%%%%%%%%%%%%%%%%%%%%%%%%%%
\begin{figure}[htb]
\epsfxsize=7cm
\centerline{\epsfbox{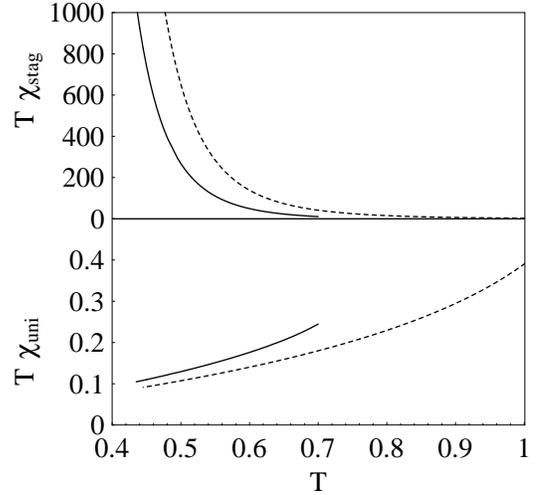}} 
\vspace{0.1cm}
\caption{Plots of  $T\chi_{\rm stag}$ for the 
staggered susceptibility and
$T \chi_{\rm uni}$ for the uniform susceptibility  
as a function of the temperature $T$:  the solid lines for
 $J_{1}=J_{2}=1/2,J_{3}=1$ and  the dashed lines 
 for $J_{1}=1, J_{2}=J_{3}=1/2$ 
}
\label{fig:half4}%-----------------------------------------
\end{figure}
%%%%%%%%%%%%%%%%%%%%%%%%%%%%%%%%%%%%%%%%%%%%%
%%%%%%%%%%%%%%%%%%%%%%%%%%%%%%%%%%%%%%%%%%%%%
In Fig. \ref{fig:half4},
we plot the effective Curie constants $T \chi_{\rm uni}$ and 
$T \chi_{\rm stag}$ as a function of $T$.
It is seen that the effective Curie constant for the uniform sector
is gradually decreased as the temperature is decreased, while 
that for the staggered sector diverges, 
implying that the antiferromagnetic correlation is enhanced 
at low temperatures. 
%%%%%%%%%%%%%%%%%%%%%%%%%%%%%%%%%%%%%%%%%%%%%
%\begin{figure}[htb]
%\epsfxsize=7cm
%\centerline{\epsfbox{Spa.eps}} 
%\vspace{0.1cm}
%\caption{The specific heat $C$ per unit cell as a function of the
%temperature $T$: 
%the solid lines for
% $J_{1}=J_{2}=1/2,J_{3}=1$ and the dashed lines 
% for $J_{1}=1, J_{2}=J_{3}=1/2$ }
%\label{fig:half5}%-----------------------------------------
%\end{figure}
%%%%%%%%%%%%%%%%%%%%%%%%%%%%%%%%%%%%%%%%%%%%%
%%%%%%%%%%%%%%%%%%%%%%%%%%%%%%%%%%%%%%%%%
%%%%% Revised   DROPPED SPECIFIC HEAT 
%%%%%%%%%%%%%%%%%%%%%%%%%%%%%%%%%%%%%%%%%%%%%
%The computed results for 
%the specific heat $C$  are shown in Fig. \ref{fig:half5}.
%It is expected that the magnetic double structure may
%be observed  in the temperature-dependent specific heat:
%i.e. the low-temperature behavior is controlled by 
% the gapless magnons, while the Schottky-type hump 
%may appear in higher temperatures.
%Although the rapid increase of the specific heat in higher 
%temperatures in Fig. \ref{fig:half5} may be assigned as  
%the tendency to the Schottky-type behavior, we cannot  
%observe the  hump structure  explicitly,
%since our approach based on the SBMFT breaks down 
%at higher temperatures.
%To see such a double structure
%in the specific heat correctly, it is thus needed to improve our 
%SBMFT in the high temperature region.
%%%%%%%%%%%%%%%%%%%%%%%%%%%%%%%%%%%%%%%%%%%%%
%%%%%%%%%%%%%%%%%%%%%%%%%%%%%%%%%%%%%%%%%%%%%

As seen from the above results, the thermodynamic properties 
at finite temperatures show 
quite similar behavior both for the Haldane chain 
in a staggered field and for the quasi-1D ferrimagnetic chain, 
reflecting the magnetic double structure inherent in these 
mixed-spin systems.

%%%%%%%%%%%%%%%%%%%%%%%%%%%%%%%%%%%%%%%%%%%%%%%%%%%%%%%%%%%%%%%%%%%%%%
\section{summary}%%%%%%%%%%%%%%%%%%%%%%%%%%%%%%%%%%%%%%%%%%%%%%%%%%%%%
%%%%%%%%%%%%%%%%%%%%%%%%%%%%%%%%%%%%%%%%%%%%%%%%%%%%%%%%%%%%%%%%%%%%%%
We have studied the 2D mixed-spin model for which 
the  $s=1/2$ and $s=1$ spin chains are stacked alternately.
This mixed-spin model includes two interesting spin systems addressed recently:
the Haldane gap system in a staggered field as well as the ferrimagnetic chain.
By calculating the dispersion relation and the thermodynamic quantities
by means of the Schwinger-boson mean-field theory, we have 
discussed the magnetic double structure inherent in our mixed-spin
systems.  
%%%%%%%%%%%%%%%%%%%%%%%%%%%%%%%%%%%%%%%%%%%%%%%%%%%%%%%%%%%%%%%%%%%%%%
%           Revised,  ADIABATICALLY                                             %%%%%%%%%%%%%%%%%%%%%%%%%%%%%%%%%%%%%%%%%%%%%%%%%%%%%%%%%%%%%%%%%%%%%%
In particular, we have treated  systematically 
the quasi-1D Haldane gap system in a staggered magnetic field and also 
the  mixed-spin chain with the ferrimagnetic ground state.
This implies that these two spin systems, which have 
been studied in different contexts  experimentally, 
should possess interesting physics common to the mixed spin systems.
%%%%%%%%%%%%%%%%%%%%%%%%%%%%%%%%%%%%%%%%%%%%%%%%%%%%%%%%%%%%%%%%%%%%%%
%In particular, 
%we have shown that the Haldane gap system in a staggered field is 
%adiabatically connected  to the ferrimagnetic chain without 
%any phase transitions between them.
%This implies that these two interesting spin systems, which have 
%been studied in different contexts  experimentally, 
%should possess essentially  the same physics.
%%%%%%%%%%%%%%%%%%%%%%%%%%%%%%%%%%%%%%
We have also found that the results obtained for the staggered-field
effect on the Haldane gap system
are qualitatively consistent with the experimental findings
in the rare-earth compounds $R_2 \rm BaNiO_{5}$.
It remains an interesting problem to evaluate dynamical quantities
related to the neutron scattering, etc.  Also, it is important to 
study how the string-order parameter behaves,\cite{Stri} when the 
system changes from the Haldane system to the ferrimagnetic chain.
These problems are now under consideration.

%%%%%%%%%%%%%%%%%%%%%%%%%%%%%%%%%%%%%%%%%%%%%%%%%%%%%%%%%%%%%%%%%%%%%%
\section*{acknowledgments}%%%%%%%%%%%%%%%%%%%%%%%%%%%%%%%%%%%%%%%%%%%%
%%%%%%%%%%%%%%%%%%%%%%%%%%%%%%%%%%%%%%%%%%%%%%%%%%%%%%%%%%%%%%%%%%%%%%
The work is partly supported by a 
Grant-in-Aid from the Ministry of Education, Science, Sports, and Culture.
A. K. is supported by the Japan Society for the Promotion of Science.
N. K. wishes to thank K. Ueda for the warm hospitality at ISSP, 
 University of Tokyo.

%%%%%%%%%%%%%%%%%%%%%%%%%%%%%%%%%%%%%%%%%%%%%%%%%%%%%%%%%%%%%%%%%%%%%%

%\end{multicols}


\begin{thebibliography}{99}
\bibitem{SWT}
There is another approach based on the modified spin wave theory,
which is similar to the SBMFT in many respects. For mixed-spin systems, 
however, the latter possesses some ambiguity for imposing the constraint 
to incorporate non-linear effects (see ref. 16), whereas the 
constraint in the SBMFT is unambiguously introduced by preserving the 
number of bosons at each site, making the SBMFT more tractable in 
our case.
\end{thebibliography}
\end{document}